\documentstyle[12pt,epsfig,cite]{article}
\newcommand{\ba}{\begin{array}}
\newcommand{\ea}{\end{array}}
\newcommand{\be}{\begin{equation}}
\newcommand{\ee}{\end{equation}}
\newcommand{\bea}{\begin{eqnarray}}
\newcommand{\eea}{\end{eqnarray}}

\newcommand{\la}{\langle}
\newcommand{\ra}{\rangle}

\newcommand{\p}{\partial}

\newcommand{\SL}[0]{SL(2,R)}

\def\CF{{\mathcal{F}}}
\def\CP{{\mathcal{P}}}
\def\CQ{{\mathcal{Q}}}
\def\IB{\relax\hbox{$\inbar\kern-.3em{\rm B}$}}
\def\IC{\relax\hbox{$\inbar\kern-.3em{\rm C}$}}
\def\ID{\relax\hbox{$\inbar\kern-.3em{\rm D}$}}
\def\IE{\relax\hbox{$\inbar\kern-.3em{\rm E}$}}
\def\IF{\relax\hbox{$\inbar\kern-.3em{\rm F}$}}
\def\IG{\relax\hbox{$\inbar\kern-.3em{\rm G}$}}
\def\IGa{\relax\hbox{${\rm I}\kern-.18em\Gamma$}}
\def\IH{\relax{\rm I\kern-.18em H}}
\def\IK{\relax{\rm I\kern-.18em K}}
\def\IL{\relax{\rm I\kern-.18em L}}
\def\IP{\relax{\rm I\kern-.18em P}}
\def\IR{\relax{\rm I\kern-.18em R}}
\def\IZ{\relax{\rm Z\kern-.5em Z}}

\def\half{\frac{1}{2}}
\def\p{\partial}
\def\f{\frac}

\begin{document}

\begin{titlepage}

\rightline{OUTP-01-38-P}
\rightline{July 2001}
\rightline{hep-th/0107160}

\vskip 2 cm

\begin{center}
{\LARGE  $SU(2)_0$ and $OSp(2|2)_{-2}$ WZNW models : Two current algebras, one Logarithmic CFT}
\vskip 1 cm

{\large I.I.Kogan\footnote{i.kogan@physics.ox.ac.uk} ~and A. Nichols\footnote{a.nichols1@physics.ox.ac.uk}}

\vskip 1 cm

{\it Theoretical Physics, Department of Physics, University of Oxford}\\
{\it 1 Keble Road, Oxford, OX1 3NP,  UK}

\vskip .5 cm 

\end{center}

\begin{abstract}

We show that the $SU(2)_0$ WZNW model has a hidden $OSp(2|2)_{-2}$ symmetry. Both these theories are known to have logarithms in their correlation functions. We also show that, like $OSp(2|2)_{-2}$, the logarithmic structure present in the $SU(2)_0$ model is due to the underlying $c=-2$ sector. We demonstrate that the quantum Hamiltonian reduction of $SU(2)_0$ leads very directly to the correlation functions of the $c=-2$ model. We also discuss some of the novel boundary effects which can take place in this model.

\end{abstract}

\end{titlepage}

\newpage

\section{Introduction}

The study of conformal invariance in two dimensions has been an extremely interesting and fruitful area of research for the last twenty years \cite{Belavin:1984vu}. 

During the last ten years an interesting class of conformal field theories (CFTs) has emerged called logarithmic conformal field theories (LCFTs). Logarithmic singularities in correlation functions were first found in \cite{Rozansky:1992rx}. In \cite{Gurarie:1993xq} the concept of LCFT was introduced and the presence of logarithmic structure was explained by the indecomposable representations that can occur in the fusion of primary operators. These occur when there are fields with degenerate scaling dimensions having a Jordan block structure.

LCFTs have emerged in many different areas such as: WNZW models and gravitational dressing \cite{Bilal:1994nx,Caux:1997kq,Giribet:2001qq}, polymers \cite{Saleur:1992hk,Cardy,Gurarie:1999yx}, disordered systems and Quantum Hall effect \cite{Caux:1996nm,Kogan:1996wk,Maassarani:1996jn,CTT,Caux:1998eu,Bhaseen:1999nm,Kogan:1999hz,Gurarie:1999bp,BernardLC,Bhaseen:2000bm,Bhaseen:2000mi,Ludwig:2000em}, string theory \cite{Kogan:1996zv,Kogan:1996df,Kogan:1999bn,Kogan:1999hz,Nichols:2000mk,Kogan:2000nw,Moghimi-Araghi:2001fg} and 2d turbulence \cite{Flohr:1996ik}.

There has also been much work on analysing the general structure and consistency of such models in particular the $c_{p,q}$ models and the special case of $c=-2$ which is by far the best understood \cite{Kausch:1995py,Gaberdiel:1998ps,Kausch:2000fu,Flohr:2000mc}. It is unclear as yet how much of the structure, for instance the role of extended algebras, is generic to all LCFTs. For more about the general structure of LCFT see \cite{Rohsiepe:1996qj,Kogan:1997fd} and references therein.

One of next best known LCFTs beyond the minimal models is the $SU(2)_0$ theory. It is a simple example of an LCFT in which we have an extended Kac-Moody symmetry. For another recent example of LCFT based on an $SU(2)$ WZNW model (at fractional level) see \cite{Gaberdiel:2001ny}. The logarithmic operators present its spectrum have been discussed previously in the context of both string theory and condensed matter \cite{Caux:1997kq,Kogan:1996df,Nichols:2001du}. The $SU(2)_0$ model is also the first studied example of an LCFT at $c=0$ which is perhaps a special sub-class of LCFTs. These models are of utmost importance for both disordered systems and critical strings \cite{Cardy,Gurarie:1999bp,Gurarie:1999yx,Moghimi-Araghi:2000dt,Kogan:2000nw}. We shall not discuss the structure of the stress tensor and its logarithmic partners here \cite{WorkinProgress}.

We shall show that this model has an extended non-local $OSp(2|2)_{-2}$ Kac-Moody symmetry in addition to the $SU(2)_0$ affine Lie algebra. The appearance of a hidden symmetry present in LCFTs was conjectured some time ago \cite{Caux:1996nm,Caux:1997kq}. We shall discuss several correspondences between this model and the well studied $c=-2$ model. The similarity of the conformal blocks has been previously noted in \cite{Bhaseen:2000mi}.

A connection between BRST  and singletons was noted in \cite{Kogan:1999bn}. Recently the appearance of a new type of BRST symmetry in LCFT was discussed \cite{Moghimi-Araghi:2001fg}. It is unclear at present if these are related or not and what the importance of them is. We also found a BRST structure which was due to the underlying topological nature of the theory.

\section{Lagrangian description of $SU(2)_0$}

Using a Lagrangian approach for $SU(2)_k$ we can obtain a free field, or Wakimoto, representation \cite{Wakimoto:1986gf}. Here we shall briefly repeat the description as the topological nature of the theory \cite{Yoshii:1992qz} becomes particularly evident. We shall follow the presentation given in \cite{Gerasimov:1990fi}. 

The classical action is just the normal sigma model:
\be \label{eqn:classL}
S_{WZNW}=\f{k}{8 \pi} \int{ d^2 z Tr g^{-1} \p^{\mu}g g^{-1} \p_{\mu} g} + \f{ik}{12 \pi} \int{ d^3 z Tr g^{-1}dg g^{-1}dg g^{-1}dg}
\ee
Clearly the case $k=0$ is quite special as this classical action vanishes.
Using the Gauss decomposition:
\be
g= \left( \begin{array}{ll} 1 & \Psi  \\ 0 & 1 \end{array} \right)
\left( \ba{ll} e^{\Phi} & 0 \\ 0 & e^{-\Phi} \ea \right)
\left( \ba{ll} 1 & 0 \\ \chi & 0 \ea \right)
\ee
In the following unless otherwise stated we restrict attention to the holomorphic sector as the anti-holomorphic one behaves in a similar way.
The classical conserved currents are:
\be
k g^{-1}dg=k \left( \ba{ll} w \chi + d \Phi & w \\ -w \chi^2 -2\chi d\Phi + d\chi & -w\chi -d\Phi\ea \right)= J^a \sigma^a
\ee
\be \label{eqn:W}
W=kw=ke^{-2\Phi}d \Psi
\ee
With this field redefinition the Lagrangian becomes:
\be
{\mathcal{L}}=-\f{1}{4\pi} (W \bar{\p}\chi + k \p{\Phi}\bar{\p}\Phi )
\ee
So far the results are purely classical however in the full path integral the transformation (\ref{eqn:W}) is anomalous and one must take into account the change in the measure. The full quantum action becomes:
\be
{\mathcal{L}}_q= -\f{1}{4 \pi} (W \bar{\p}\chi+ (k+2) \p \Phi \bar{\p} \Phi + {\mathcal{R}} \Phi )
\ee
To get the standard normalisation we rescale $ -(k+2) \p \Phi \bar{\p} \Phi = \f{1}{2} \p \phi \bar{\p} \phi $. We also redefine $W=-\beta~,~\chi=\gamma$. Then :
\be
{\mathcal{L}}_q= -\f{1}{4 \pi} (-\beta \bar{\p} \gamma -\half \p \phi \bar{\p} \phi + {\mathcal{R}} \Phi )
\ee
The stress tensor then becomes:
\be \label{eqn:stressT}
T=-\beta \p \gamma -\half \p \phi \p \phi - \f{i}{\sqrt{2(k+2)}} \p^2 \phi
\ee
We observe that the stress tensor is composed of two commuting parts: the $\beta \gamma$ system with $c=2$ and for $k=0$ the $\phi$ system has $c=-2$. This latter system is the same as the bosonised symplectic fermion system:
\bea
&&\xi \sim e^{i \phi} ~~~~~~~ \eta \sim e^{-i \phi} ~~~~ i \p \phi =  \xi \eta ~~~ \xi(z)\eta(w) \sim \f{1}{z-w} \sim \eta(z)\xi(w) 
\eea
Although the classical action  (\ref{eqn:classL}) vanished the stress tensor is non-trivial due to the transformation of the measure. This is a hallmark of a topological field theory \cite{Witten:1988ze}. We can see this explicitly by considering the nilpotent BRST dimension one operator $Q=\beta \eta$. $Q$ induces the following transformations on the fields:
\be
\delta \xi=\beta ~~~~ \delta \beta=0 ~~~~ \delta \gamma=-\eta ~~~~ \delta \eta=0
\ee
The currents and stress tensor become BRST exact $J^a=\delta \Phi^a ~~~ T=\delta G$. As this is a topological model it can also easily be written as a twisted ${\mathcal{N}}=2$ theory where the previous BRST charge $Q$ is the zero mode of the field $G_+$:
\bea
G_+&=& \beta \eta \nonumber \\
G_-&=&-\xi \p \gamma \\
J&=&\xi \eta \nonumber \\
T&=&-\beta \p \gamma - \xi \p \eta \nonumber
\eea
In order to get a non-trivial conformal field theory we are going to extend (by hand at the moment) the structure by assuming that the stress tensor, $T$, and $SU(2)$ affine currents, $J^a$, do not decouple. The theory is then non-topological in the sense that correlation functions gain a non-trivial coordinate dependence.
\section{Emergence of $OSp(2|2)_{-2}$ from the free field representation of $SU(2)_0$}
We now return to using the $\beta,\gamma$ system.
For $k=0$ the stress tensor (\ref{eqn:stressT}) has $SU(2)_0$ symmetry generated by the currents:
\be
J^+=\beta ~~~~~~
J^3=\xi \eta +\gamma \beta ~~~~~~
J^-=-2 \xi \eta \gamma - \beta\gamma^2
\ee
where $\beta$ and $\gamma$ obey the standard free field relations:
\be
\beta(z) \beta(w) \sim 0 \sim \gamma(z) \gamma(w) ~~~~ \beta(z) \gamma(w) \sim \f{-1}{z-w} \sim -\gamma(z) \beta(w)
\ee
these obey the standard Kac-Moody algebra with zero central extension:
\be \label{eqn:SU2KM}
J^3(z) J^{\pm}(w) \sim \pm \f{J^{\pm}(w)}{z-w} ~~~ J^+(z)J^-(w) \sim \f{2 J^3(w)}{z-w} ~~~ J^3(z)J^3(w) \sim 0
\ee
The stress tensor is just the standard Sugawara one (See Appendix):
\bea \label{eqn:sugawara}
T&=&\f{1}{2}\left( \half J^+J^- +\half J^-J^+ + J^3J^3 \right) 
= -\beta \p \gamma - \xi \p \eta
\eea
We can also express the $(\beta,\gamma)$ part in terms of free bosons:
\be
\beta=\f{1}{\sqrt{2}}(\p \phi_1 + \p \phi_2) ~~~~ \gamma=\f{1}{\sqrt{2}}( \phi_1- \phi_2)
\ee
In this way the stress tensor becomes:
\be \label{eqn:stressOSp}
T=-\half \p \phi_1 \p \phi_1 + \half \p \phi_2 \p \phi_2 -\xi \p \eta
\ee
and the currents are:
\bea \label{eqn:newSU2KM}
J^+&=&\f{1}{\sqrt{2}}(\p \phi_1 + \p \phi_2) \nonumber \\
J^3&=&\xi \eta + \f{1}{2}( \phi_1- \phi_2) (\p \phi_1 + \p \phi_2) \\
J^-&=&- \sqrt{2} \xi \eta ( \phi_1- \phi_2) - \f{1}{2 \sqrt{2}}(\p \phi_1 + \p \phi_2)( \phi_1- \phi_2)^2 \nonumber
\eea
Note that the bosons naturally come in a pair; one compact and the other non-compact with the OPEs:
\be
\phi_1(z) \phi_1(w) \sim -\ln(z-w) ~~~~ \phi_2(z) \phi_2(w) \sim \ln(z-w) ~~~~ \phi_1(z) \phi_2(w) \sim 0
\ee
It is natural now to consider using the fields to also create fermionic generators. Indeed it is known from \cite{Bhaseen:2000mi,Ludwig:2000em} that the stress tensor (\ref{eqn:stressOSp}) is exactly the one given by the Sugawara construction of $OSp(2|2)_{-2}$.
The currents of $OSp(2|2)_{-2}$ are:
\bea \label{eqn:OSpcurrents}
H&=&-i \sqrt{2} \p \phi_1 ~~~~~~~~~~~~~~~~
J=-i \sqrt{2} \p \phi_2 \nonumber \\
K&=&2 e^{-i \sqrt{2} \phi_1} ~~~~~~~~~~~~~~~~~
\hat{K}=-2 e^{i \sqrt{2} \phi_1} \nonumber \\
\hat{G}_+&=& \sqrt{2} e^{\f{i}{\sqrt{2}}(\phi_1+\phi_2)} \p \eta ~~~~~~
G_+=-\sqrt{2} e^{\f{i}{\sqrt{2}}(-\phi_1+\phi_2)} \p \eta \\
\hat{G}_-&=&-\sqrt{2} e^{\f{i}{\sqrt{2}}(\phi_1-\phi_2)} \xi ~~~~~~
G_-= \sqrt{2} e^{\f{i}{\sqrt{2}}(-\phi_1-\phi_2)} \xi \nonumber 
\eea
These obey the $OSp(2|2)$ algebra at level $k=-2$ (we use the notation of \cite{Maassarani:1996jn}): 
\bea
J(z) J(w) \sim \f{k}{(z-w)^2} ~~~~ H(z) H(w) \sim \f{-k}{(z-w)^2} \nonumber \\
J(z) G_{\pm}(w) \sim \pm \f{G_{\pm}(w)}{z-w} ~~~~ J(z) \hat{G}_{\pm}(w) \sim \pm \f{\hat{G}_{\pm}(w)}{z-w} \nonumber \\
H(z) G_{\pm}(w) \sim \f{G_{\pm}(w)}{z-w} ~~~~ H(z) \hat{G}_{\pm}(w) \sim - \f{\hat{G}_{\pm}(w)}{z-w} \nonumber \\
H(z) K(w) \sim \f{2 K(w)}{z-w} ~~~~ H(z) \hat{K}(w) \sim \f{-2 \hat{K}(w)}{z-w}\nonumber \\
\hat{G}_{\pm}(z) G_{\mp}(w) \sim \f{k}{(z-w)^2} + \f{H(w) \pm J(w)}{z-w} \\
\hat{K}(z) K(w) \sim \f{2k}{(z-w)^2} + \f{4 H(w)}{z-w} \nonumber \\
G_-(z) G_+(w) \sim \f{K(w)}{z-w} ~~~~
\hat{G}_-(z) \hat{G}_+(w) \sim \f{\hat{K}(w)}{z-w} \nonumber \\
K(z) \hat{G}_{\pm}(w) \sim \f{-2 G_{\pm}}{z-w} ~~~~
\hat{K}(z) G_{\pm}(w) \sim \f{2 \hat{G}_{\pm}(w)}{z-w}\nonumber
\eea
Clearly there are many other conserved spin one currents one may construct from these free fields. It is possible to construct ones which also form a closed algebra. However as this is an affine algebra we can also form the Sugawara tensor from the above currents.

The Sugawara tensor for general $OSp(2|2)_k$ is:
\be \label{eqn:OSpsugawara}
T=\f{1}{4-2k} \left( HH-JJ -\half(K \hat{K} + \hat{K} K) + \hat{G}_+ G_- - G_- \hat{G}_+ + \hat{G}_- G_+ - G_+ \hat{G}_- \right)
\ee
Using the expressions for the currents (\ref{eqn:OSpcurrents}) in the case $k=-2$ we find this gives exactly the same stress tensor (\ref{eqn:stressOSp}) (see Appendix) as that of the $SU(2)_0$ theory.

The only common operator between the $OSp(2|2)_{-2}$ and $SU(2)_0$ (\ref{eqn:newSU2KM}) algebras is $J^+=\f{i}{2}(J+H)$. For the other operators we get more complicated expressions for example:
\bea
J(z) J^3(w) &=& -i \sqrt{2} \p \phi_2(z) ~~ \left( \xi \eta + \f{1}{2}( \phi_1- \phi_2) (\p \phi_1 + \p \phi_2) \right) (w) \nonumber \\
& \sim & \f{i \gamma(w)}{(z-w)^2} + \f{i J^+(w)}{z-w}
\eea
Thus combining these two algebras together clearly does not produce a normal affine Lie algebra as we have non-trivial operators of dimension zero in the OPE. We have not determined what the overall algebra is but it seems to naturally have indecomposable representations and logarithmic terms in it.

\subsection{Extra $c=-2$ structure}
\label{sec:secondc=-2}
There is another $c=-2$ structure which arises through bosonising the $c=2$ bosonic ghost system \cite{Friedan:1986ge}:
\be
\beta=e^{-\psi} \p \xi_1 ~~~~~~~~~~~~ \gamma=e^{\psi} \eta_1
\ee
The bosonic $\psi$ system has $c=4$ with the stress tensor:
\be
T_{\psi}=-\half \p \psi \p \psi - \half \p^2 \psi
\ee
The fermionic $\xi_1,\eta_1$ is \emph{another} $c=-2$ system with conformal weights $0$ and $1$ respectively. It is however essential to note that due to the appearance of $\p \xi_1$ the zero mode is \emph{not} present. This means that one expects there to be no logarithmic solutions in this part. To write the system entirely in terms of bosons we can repeat as before:
\be
\eta_1=e^{-\chi} ~~~~~~~~ \xi=e^{\chi}
\ee
Then the full bosonised expressions are:
\be
\beta=e^{-\psi+\chi} \p \chi ~~~~~~~~ \gamma=e^{\psi-\chi}
\ee
The total stress tensor for the full $SU(2)_0$ theory can now be written purely in terms of bosonic fields:
\be
T= - \half \p \psi \p \psi + \half \p \phi \p \phi + \half \p \chi \p \chi - \half \p^2 \psi + \half \p^2 \chi  + \half \p^2 \phi
\ee
We thus see a very peculiar symmetry that arises in $SU(2)_0$ namely a $Z_2$ symmetry exchanging $\chi$ and $\phi$ fields. The kinetic terms admit a continuous rotation between the two fields but the $ \p^2 \chi  +  \p^2 \phi$ breaks this to $Z_2$. This symmetry is exchanging the two $c=-2$ subsystems.
There is also an interesting connection between $OSp(2|2)_{-2}, SU(2)_0$ and the model $OSp(4|4)_1$. This is based on the observation \cite{BernardLC}:
\be
T_{OSp(4|4)_1}=T_{OSp(2|2)_{-2}}+T_{SU(2)_0}
\ee
The $c=2$ sector of $SU(2)_0$ when written as $c=-2$ and $c=4$ parts gives the overall structure $c=4+(-2-2)=0$ which is precisely the same of $OSp(4|4)_1$.
The $OSp(2|2)_{-2}$ and $SU(2)_0$ models as we shall see have logarithmic operators in their spectrum. However $OSp(4|4)_1$ has a free field representation with non-logarithmic correlation functions. However despite this, for certain correlators, it is possible to braid these logarithmic conformal blocks into an $OSp(4|4)_1$ correlator which does not contain logarithmic terms \cite{Bhaseen:2000bm}. 

The correlation functions did not reveal the presence of any negative dimension operators in the fusion of the primaries. This fact and the assumption that any logarithmic partner of the identity should be an $SU(2)$ singlet actually shows that the vacuum is unique.\footnote{We thank A. Lewis for pointing this out to us.}

To see this we use the expression for $L_{0}$ derived from (\ref{eqn:sugawara}):
\bea
L_0=\f{1}{2} \sum_{n=-\infty}^{+\infty} :J^a_{-n}J^a_{n}:
\eea
Now acting on a state $\left. |\omega \right>$ at $h=0$ we see that if we wish it to be an $SU(2)$ singlet then $\left. J^a_0|\omega \right>=0$:
\bea
L_0\left. |\omega \right>=\f{1}{2} \sum_{n=1}^{\infty}{J^a_{-n}}J^a_n \left. |\omega \right>
\eea

Clearly if there are no negative dimension operators in the theory then we must have $J^a_n\left. |\omega \right> =0$ for $n > 0$ otherwise this would generate states with $h < 0$. Thus we get $L_0 \left. |\omega \right>  =0$. It is in particular not possible to have a logarithmic partner of the normal identity ($\left. |\Omega \right>$) with $L_{0} \left. |\omega \right> =\left. |\Omega \right>$.

At this stage it is not completely clear how the degenerate vacua of the two $c=-2$ sub-theories are combined to produce such a single state and here we shall only suggest a possibility. In addition to the two normal vacua, $\left. |\Omega_i \right>$ (where $i=1,2$ labels the two $c=-2$ sub-theories) we have their logarithmic partners $\left. |\omega_i \right>$. They satisfy:
\bea \label{eqn:vacua}
L^i_0 \left. |\omega_i \right>=\left. |\Omega_i \right> ~~~~~ L^i_0 \left. |\Omega_i \right>=0
\eea
where $L^i_n$ is the mode expansion of the stress tensor in each of the two sub-theories. It is now possible to combine these with a symplectic structure to create a single non-degenerate vacuum:
\bea
\left. |0 \right>=\left. |\omega_1 \right>\left. |\Omega_2 \right> - \left. |\Omega_1 \right> \left. |\omega_2 \right>
\eea
Using (\ref{eqn:vacua}) we see that this satisfies:
\bea
L_0 \left. |0 \right>=(L^1_0+L^2_0) \left. |0 \right>&=&(L^1_0+L^2_0) \left( \left. |\omega_1 \right>\left. |\Omega_2 \right> - \left. |\Omega_1 \right> \left. |\omega_2 \right> \right) \\
&=& \left. |\Omega_1 \right>\left. |\Omega_2 \right> - \left. |\Omega_1 \right>\left. |\Omega_2 \right> =0 \nonumber
\eea
In this way it is possible to combine the two logarithmic vacua to create a single non-degenerate one. If we also define the bra-state:
\bea \label{eqn:bra}
\left< 0| \right.=-\left< \omega_1| \right. \left< \Omega_2 |\right. + \left< \Omega_1| \right. \left< \omega_2 |\right.
\eea
Then by using $\left< \Omega_i|\Omega_j \right>=0$ we see that $\left< 0| \right>$ has positive norm \footnote{To get $\left.|0\right>$ to have positive norm one actually has to introduce an extra (-1) in (\ref{eqn:bra}) in the definition of the bra-state. We do not know how to justify this extra phase.}:
\bea
\left<0|0 \right>=2 \left< \omega_1| \Omega_1 \right>\left< \Omega_2| \omega_2 \right> \ne 0
\eea
We have discussed here only the production of a chiral vacuum. It is actually non-trivial even in $c=-2$ to produce a local logarithmic vacuum sector from the chiral states \cite{Gaberdiel:1998ps}.

\section{Correlation functions from KZ equations}

\label{sec:2and3pt}

It will be convenient in much of this paper to introduce the following representation for the  $SU(2)$ generators \cite{Zamolodchikov:1986bd}:
\bea \label{eqn:repn}
J^+=x^2\frac{\p}{\p x}-2jx, ~~~ 
J^-=-\frac{\p}{\p x}, ~~~
J^3=x\frac{\p}{\p x}-j \eea
There is also a similar algebra in terms of $\bar{x}$ for the antiholomorphic part. It is easily verified that these obey the global $SU(2)$ algebra.

We introduce primary fields, $\phi_j(x,z)$  of the KM algebra:
\be
J^a(z)\phi_j(x,w) = \frac{1}{z-w} J^a(x) \phi_j(x,w) 
\ee
where $J^a(x)$ is given by (\ref{eqn:repn}). The fields $\phi_j(x,z)$ are also primary with respect to the Virasoro algebra with $L_0$ eigenvalue:
\be
h=\frac{j(j+1)}{k+2}
\ee
The two point functions are fully determined using global $SU(2)$ and conformal transformations and can be normalised in the standard way:
\be
\la \phi_{j_1}(x_1,z_1) \phi_{j_2}(x_2,z_2) \ra = \delta^{j_1}_{j_2}x_{12}^{2j_1}z_{12}^{-2h}
\ee
The general form of the three point function is:
\bea
\la \phi_{j_1}(x_1,z_1) \phi_{j_2}(x_2,z_2) \phi_{j_3}(x_3,z_3) \ra = C(j_1,j_2,j_3)~~ x_{12}^{j_1+j_2-j_3} x_{13}^{j_1+j_3-j_2} x_{23}^{j_2+j_3-j_1} \\
z_{12}^{-h_1-h_2+h_3} z_{13}^{-h_1-h_3+h_2} z_{23}^{-h_2-h_3+h_1} \nonumber
\eea
The $C(j_1,j_2,j_3)$ are the structure constants which in principle completely determine the entire theory.

For the case of the four point correlation functions of $SU(2)$ primaries the form is determined by global conformal and $SU(2)$ transformations up to a function of the cross ratios.
\bea \label{eqn:correl} 
\langle \phi_{j_1}(x_1,z_1) \phi_{j_2}(x_2,z_2) \phi_{j_3}(x_3,z_3) \phi_{j_4}(x_4,z_4) \rangle
&=&z_{43}^{h_2+h_1-h_4-h_3}z_{42}^{-2h_2}z_{41}^{h_3+h_2-h_4-h_1} \nonumber \\
& & z_{31}^{h_4-h_1-h_2-h_3}x_{43}^{-j_2-j_1+j_4+j_3}x_{42}^{2j_2}  \\
& & x_{41}^{-j_3-j_2+j_4+j_1}x_{31}^{-j_4+j_1+j_2+j_3}~F(x,z) \nonumber
\eea 
Here the invariant cross ratios are:
\be 
x=\frac{x_{21}x_{43}}{x_{31}x_{42}} ~~~ z=\frac{z_{21}z_{43}}{z_{31}z_{42}} 
\ee
\subsection{The Knizhnik-Zamolodchikov Equation}

Correlation functions of the WZNW model satisfy a set of partial differential equations known as  Knizhnik-Zamolodchikov (KZ) equation due to constraints from the null states following from (\ref{eqn:sugawara}). These are:
\be
|\chi \ra = ( L_{-1} - \frac{1}{k+2} \eta_{ab}J^a_{-1}J^b_0 ) |\phi \ra
\ee
For two and three point functions this gives no new information. However for the four point function (\ref{eqn:correl}) it becomes a partial differential equation for $F(x,z)$. For  a compact Lie group this equation can be solved \cite {Knizhnik:1984nr} as it reduces to a set of ordinary differential equations.
\be
\left[(k+2) \frac{\p}{\p z_i}+\sum_{j\neq i}\frac{\eta_{ab} J^a_i \otimes J^b_j}{z_i-z_j} \right] \left<\phi_{j_1}(z_1) \cdots \phi_{j_n}(z_n) \right> =0 
\ee
where $k$ is the level of $SU(2)$ WZNW model. 

If we now use our representation (\ref{eqn:repn}) we find the KZ equation for four point functions is:
\be \label{eqn:KZ}
(k+2) \frac{\p}{\p z} F(x,z)=\left[ \frac{\CP}{z}+\frac{\CQ}{z-1} \right] F(x,z)
\ee
Explicitly these are:
\bea
\CP \!\!\!\!&=&\!\!-x^2(1-x)\frac{\p^2}{\p x^2}+((-j_1-j_2-j_3+j_4+1)x^2+2j_1x+2j_2x(1-x))\frac{\p}{\p x} \nonumber \\
& & +2j_2(j_1+j_2+j_3-j_4)x-2j_1j_2 \\
\CQ \!\!\!\!&=&\!\!-(1-x)^2x\frac{\p^2}{\p x^2}-((-j_1-j_2-j_3+j_4+1)(1-x)^2+2j_3(1-x)+2j_2x(1-x))\frac{\p}{\p x} \nonumber \\
& & +2j_2(j_1+j_2+j_3-j_4)(1-x)-2j_2j_3 
\eea
\subsection{Correlation functions}
The correlation functions of the fundamental $j=\half$ operators in $SU(2)_0$ were found in \cite{Caux:1997kq}. Using the auxiliary variable $x$ they can be written as:
\bea \label{eqn:confblocks}
\CF_1(x,z)&=& z^{\f{1}{4}}(1-z)^{\f{1}{4}} \left\{ \left( -\f{E}{z(1-z)}+\f{K}{z} \right) x + \f{E}{1-z} \right\} \\
\CF_2(x,z)&=& z^{\f{1}{4}}(1-z)^{\f{1}{4}} \left\{ \left( \f{\tilde{E}}{z(1-z)}-\f{\tilde{K}}{1-z} \right) x + \f{\tilde{K}}{1-z} -\f{\tilde{E}}{1-z} \right\} \nonumber
\eea
where we use the notation:
\be
E(z)={}_2F_1 \left(\half,-\half;1;z \right) ~~~ K(z)={}_2F_1 \left( \half,\half;1;z \right) ~~~ \tilde{E}(z)=E(1-z) ~~~ \tilde{K}(z)=K(1-z)
\ee
In the simple discrete representations it is more usual to rewrite these using the standard index notation. Then we get:
\bea
\left< g_{\epsilon_1,\bar{\epsilon}_1} (z_1,\bar{z}_1)  g^{\dagger}_{\bar{\epsilon}_2,\epsilon_2} (z_2,\bar{z}_2)  g_{\epsilon_3,\bar{\epsilon}_3} (z_3,\bar{z}_3)  g^{\dagger}_{\bar{\epsilon}_4,\epsilon_4} (z_4,\bar{z}_4) \right> &=& 
|z_{13}z_{24}|^{-\f{3}{2}} |z(1-z)|^{\f{1}{2}} \\
&&\sum_{i,j=1,2}I_i\bar{I}_j (F_a^i \bar{F}_b^j + F_b^i \bar{F}_a^j) \nonumber
\eea
These invariant tensors are $I_1=\delta_{\epsilon_1,\epsilon_2} \delta_{\epsilon_3,\epsilon_4} ~~~ I_2=\delta_{\epsilon_1,\epsilon_4} \delta_{\epsilon_2,\epsilon_3} $ and the functions $F_{a,b}^{i,j}$ are given by:
\bea \label{eqn:SU2blocks}
F_a^1&=&\f{\pi}{4} F(\f{1}{2},\f{3}{2};2;z)=\f{K-E}{z} ~~~~~
F_b^1=\f{\pi}{2} F(\f{1}{2},\f{3}{2};1;1-z)=\f{\tilde{E}}{z} \\
F_a^2&=& \f{\pi}{2} F(\f{1}{2},\f{3}{2};1;z)=\f{E}{1-z} ~~~~~
F_b^2=\f{\pi}{4} F(\f{1}{2},\f{3}{2};2;1-z)=\f{\tilde{K}-\tilde{E}}{1-z} \nonumber
\eea
As an example we take the correlator:
\bea
\left< g_{++}g^{\dagger}_{++}g_{--}g^{\dagger}_{--} \right> = 
|z_{13}z_{24}|^{-\f{3}{2}} |z(1-z)|^{\f{1}{2}} \left( \f{E}{1-z} \overline{\f{\tilde{K}-\tilde{E}}{1-z}}+ \f{\tilde{K}-\tilde{E}}{1-z} \overline{\f{E}{1-z}} \right)
\eea
In \cite{Caux:1997kq} the above correlation functions were analysed and the OPE was found:
\bea \label{eqn:SU2OPE}
g_{\epsilon_1,\bar{\epsilon}_1}(z_1,\bar{z}_1) g^{\dagger}_{\epsilon_2,\bar{\epsilon}_2}(z_2,\bar{z}_2) \sim |z_{12}|^{-\f{3}{2}} \left\{ z_{12} \delta_{\bar{\epsilon}_1,\bar{\epsilon}_2} t^i_{\epsilon_1,\epsilon_2} K^i(z_2) + \bar{z}_{12} \delta_{\epsilon_1,\epsilon_2} \bar{t}^i_{\bar{\epsilon}_1,\bar{\epsilon}_2} \bar{K}^i(z_2) \right. \nonumber \\
 \left. + |z_{12}|^2 t^i_{\epsilon_1,\epsilon_2} \bar{t}^j_{\bar{\epsilon}_1,\bar{\epsilon}_2} \left( D^{ij}(z_2,\bar{z}_2) + \ln|z_{12}| C^{ij}(z_2,\bar{z}_2) \right) + \cdots \right\}
\eea
The operators $K^i,\bar{K}^i,C$, and $D$ form a staggered module of the $SU(2)$ Kac-Moody algebra with the OPEs given by:
\bea
J^i(z) J^j(0) &\sim& \f{i \epsilon^{ijk} J^k}{z} \nonumber \\
J^i(z) K^j(0) &\sim& \f{i \epsilon^{ijk} K^k}{z} \\
K^i(z) K^j(0) &\sim& \f{\delta^{ij}}{z^2} + \f{i \epsilon^{ijk} N^k}{z} \nonumber \\
J^i(z) N^j(0) &\sim& \f{\delta^{ij}}{z^2} + \f{i \epsilon^{ijk} N^k}{z} \nonumber
\eea

For the $j=1$ operators the conformal blocks were found in \cite{Nichols:2001du}. We have also found explicit solutions for the $j=2,3$ cases. In all of these cases, with integer spin $j$, when one performs the conformal bootstrap one finds that the logarithmic blocks decouple.
\subsection{Factorisation of KZ equations}
One finds in many models, for certain values of parameters, there may be a reduced subspace of solutions on which one can perform the conformal bootstrap (i.e. construct a single-valued correlator obeying the appropriate crossing symmetries). In this way we find much simpler correlators than one would generically get.

From the $OSp(2|2)_{-2}$ algebra we can easily construct a $U(1|1)$ subalgebra with currents given by:
\be
j_1=H+J ~~~~~~ j_2=H-J ~~~~~~ \eta=\hat{G}_+ ~~~~~~  \bar{\eta}=G_-
\ee
where the currents $H,J,\hat{G}_+$ and $G_-$ are those given in (\ref{eqn:OSpcurrents}). We only differ from the notations of \cite{Mudry:1996zs} in the respect that their currents $J,j$ correspond to our $j_1,j_2$.  

These obey the algebra:
\bea
&&j_1(z) j_1(0) \sim 0 ~~~~ \nonumber
j_2(z) j_2(0) \sim 0 \\
&&j_1(z) \eta(0) \sim 0 ~~~~ \nonumber
j_1(z) \bar{\eta}(0) \sim 0 \\
&&j_1(z) j_2(0) \sim \f{4}{z^2} ~~~~~
j_2(z) \eta(0) \sim \f{-2 \eta}{z} \\
&&j_2(z) \bar{\eta}(0) \sim \f{2 \bar{\eta}}{z} ~~~~~
\eta(z) \bar{\eta}(0) \sim \f{-2}{z^2}+\f{j_1}{z} \nonumber
\eea
In the notation of \cite{Mudry:1996zs} this has $k=2,k_j=0$.
From this algebra one can form the $U(1|1)$ stress tensor via the Sugawara construction:
\be
T=\f{1}{2k}(j_1j_2+\eta \bar{\eta}- \bar{\eta}\eta) +\f{4-k_j}{8k^2} j_1 j_1
\ee
As mentioned in \cite{Ludwig:2000em} when we compute this using the free field expressions for the currents (\ref{eqn:OSpcurrents}) the above stress tensor evaluates to exactly the same as that from the \emph{full} $OSp(2|2)_{-2}$ algebra (\ref{eqn:OSpsugawara}) (See Appendix). We thus arrive at two different expressions for the stress tensor $T$:
\bea
T_{OSp(2|2)_{-2}}&=&\f{1}{8} \left( HH-JJ -\half(K \hat{K} + \hat{K} K)
 + \hat{G}_+ G_- + G_- \hat{G}_+ + \hat{G}_- G_+ + G_+ \hat{G}_- \right) \nonumber \\
T_{U(1|1)}&=&\f{1}{4} \left( (H+J)(H-J)+ \hat{G}_+ G_- - G_- \hat{G}_+ \right) +\f{1}{8} (J+H)(J+H)
\eea
These become equal when using the free field representation. The difference between these two thus appears as a Kac-Moody null vector (at level 2) of the model. This seems to be what is responsible in this case for the reduction of the subspace required for the bootstrap. We have also similarly analysed $OSp(2|2)$ at several other levels where factorisation was observed and found that null vectors are also present. 

It is not clear if all such factorisations of the KZ equations can be explained in such a manner. There are certainly null vectors present for the operators $j \in Z$ in $SU(2)_0$, where the correlation functions have been found to have a fairly simple form. The issue of null vectors is more delicate in a non-unitary theory because, contrary to the unitary case, null states are not forced to decouple from the spectrum.

We found an ansatz that gives a solution to the 4 point correlation function in $SU(2)_0$ when all four operators have the same spin, $j$. As these all belong to the finite dimensional representations we can write the general solution to the KZ equation (\ref{eqn:KZ}) as:
\be
\CF(x,z)=F_0(z)+xF_1(z)+x^2F_2(z)+ \cdots +x^{2j}F_{2j}(z)
\ee
We then substitute successively to get an ordinary differential equation (of order $2j$) in terms of the lowest component $F_0(z)$. We find that it always has a solution of the form:
\be
F_0(z)=z^{-j(j-1)}(1-z)^{-j(j+1)}F(j,-j;1;z)
\ee
We have checked this explicitly for $j \le 3$ but have no general proof of this fact.

One can then substitute back and by differentiations alone we get all of the other terms $F_1,\cdots,F_{2j}$. In this way we get a full solution without ever directly solving the differential equation.
On its own the above solution would not lead to a well behaved correlator but by using crossing symmetry and monodromy around $z=0,1, \infty$ one can determine a full set that does. It is not clear if this always is the unique solution satisfying the bootstrap but it certainly is in the cases with $j \le 3$.
\section{Free field approach for $SU(2)_0$}

 \label{sec:freefield}
The vertex operators of the primary fields of $SU(2)_0$ are given by:
\be \label{eqn:vertex}
\phi_{j,m}= \gamma^{j-m} e^{i j \phi}
\ee
They have $h=\f{j(j+1)}{2}$ and obey:
\bea
J^{\pm}(z)\phi_{j,m}(w) \sim (j \mp m) \frac{\phi_{j,m \pm 1}}{z-w} ~~~~ 
J^3(z)\phi_{j,m}(w) \sim  m \frac{\phi_{j,m}}{z-w} 
\eea
Here we shall mostly restrict attention to the fundamental $j=\half$ doublet having $h=\f{3}{8}$. This is exactly the conformal weight of the doublet field, $\nu^{\alpha}$ ($\alpha=\pm$), in the $c=-2$ theory. The field $\nu^{\pm}$ transforms under a global $SU(2)$ isospin symmetry that is present in the $c=-2$ theory. It should be stressed that this global $SU(2)$ symmetry rotates us between the fermionic ghosts $\xi$ and $\eta$ and is \emph{not} the same as the affine $SU(2)_0$ symmetry. The operators (\ref{eqn:vertex}) become:
\be \label{eqn:normalvertex}
V_+=\nu^+ ~~~~ V_-=\gamma \nu^+
\ee
The generators of $SU(2)_0$ do not change the isospin index of the $\nu^{\alpha}$ field they only affect the $m$ dependent $\gamma$ part. The origin of this isospin structure can be seen from the fact that the four point function is actually of the form:
\be
\left< g(z_1,\bar{z_1})  g^{\dagger}(z_2,\bar{z_2})  g(z_3,\bar{z_3})  g^{\dagger}(z_4,\bar{z_4}) \right>
\ee
We call these other fields $g^{\dagger}$ the conjugate fields; they are not the complex conjugates but have transformation under $G \in SU(2)$ as:
\be
g \rightarrow Gg ~~~~~ g^{\dagger}  \rightarrow g^{\dagger}G^{\dagger}
\ee
The conjugate fields are clearly required to ensure that the overall group invariance of the correlator. From the $c=-2$ point of view all this is obvious as the only non-vanishing correlators are isospin singlets. The isospin symmetry thus rotates between the fields and their conjugates.

In the free field representation one also introduces dual operators in order to calculate correlation functions. There are several choices for these all of which lead to the same answers. 
\begin{itemize}
\item
If we use the dual operators of Dotsenko \cite{Dotsenko:1990ui} where the highest component is given by:
\be
\tilde{V}_{j,j}=\beta^{2j-1-k}e^{-i \sqrt{\f{2}{k+2}}(j-k-1)\phi}
\ee
which also transform in the doublet representation of $SU(2)$ we find they are just the same as our original operators. 
\item The dual operators of Gerasimov et al.\cite{Gerasimov:1990fi} which do \emph{not} transform in a finite dimensional representation of $SU(2)$, instead behaving like the infinite dimensional $SU(2)$ representation $j=-\f{3}{2}$ with $h=\f{3}{8}$, have the form:
\be \label{eqn:conjugatevertex}
\tilde{V}_+=\gamma^{-2} \nu^{-} ~~~~ \tilde{V}_-=\gamma^{-1} \nu^{-}
\ee
The isospin symmetry thus also seems in this case to be relating the fields and their duals.

\end{itemize}

\subsection{Free field calculation}

We now wish to show that although the two theories appear to be decoupled from the point of view of the Lagrangian, or equivalently the stress tensor $T$, the screening charge is mixes them in a non-trivial way.
 We calculate the following simple correlator:
\be \label{eqn:examplecorrel}
<V_-(0)V_+(z)V_-(1)V_+(\infty)>_{WZW}
\ee
This correlator is expressed in terms of the original fields the WZNW theory.
In order to calculate this in the free field representation one must first insert a charge at $\infty$ that correctly reproduces the central charges of all 3 free fields. We shall denote by the subscript $s$ the correlation function that has this charge at infinity. This charge is given by:
\be
V_s(R)=e^{i \phi +u -iv}
\ee
It is the product of the charges that would be used in the $c=-2$, $c=-2$ and $c=2$ sectors. Let us also note that this is invariant under the $Z_2$ symmetry that exchanges the two $c=-2$ sectors (see section \ref{sec:secondc=-2}). In order to obtain non vanishing results the correlator must be charge neutral. In order to achieve this one introduces conjugate operators and screening charges.
We have already given expressions for the vertex operators (\ref{eqn:normalvertex}) and their conjugates (\ref{eqn:conjugatevertex}) for the case of $k=0$ using the conventions of Gerasimov et al. \cite{Gerasimov:1990fi}.

The screening charge is the integral over the dimension one field $\beta \eta$ (This is exactly the same field that occurred earlier as the BRST operator of the topological field description of $SU(2)_0$) or in terms of bosonised $u,v$ fields:
\be \label{eqn:SU2screen}
Q= \f{1}{i} \oint e^{-i \phi -u +iv}\p v
\ee
We can now evaluate the correlator (\ref{eqn:examplecorrel}):
\bea
<V_-(0)V_+(z)V_-(1)V_+(\infty)>_{WZW}&=&
<V_-(0)V_+(z)V_-(1)\tilde{V}_+(\infty) Q>_{s}  \\
&=&\oint dt <V_-(0)V_+(z)V_-(1)\tilde{V}_+(\infty) J(t)>_{s} \nonumber 
\eea
Inserting the forms of the operators given above we get
\bea
\oint dt <e^{\f{i}{2}\phi}(0)e^{\f{i}{2}\phi}(z)e^{\f{i}{2}\phi}(1)e^{\f{-3i}{2}\phi}(\infty) e^{-i \phi}(t) e^{i \phi}(R)>_{0} \nonumber \\
<e^{u-iv}(0) e^{u-iv}(1) e^{-2u+2iv}(\infty) e^{-u+iv}\p v(t) e^{u-iv}(R)>_0 \\
=\oint dt z^{\f{1}{4}} t^{-1/2} (z-1)^{-1/2} (t-z)^{-1/2} (t-1)^{-1/2} \f{2t-1}{t(t-1)} \nonumber
\eea
The subscript zero denotes that we are now explicitly writing the vacuum charge $V_s(R)$. The factor $\f{2t-1}{t(t-1)}$ is that due to the $c=2$ $\beta,\gamma$  part. In this representation one can choose two independent contours over which to perform the integral. These two solutions can be expressed in terms of hypergeometric functions and they give the same answer as the KZ equation. 

For a pure $c=-2$ theory we have the same $\phi$ dependent charge at infinity as before namely:
\be
V_s(R)=e^{i \phi}
\ee
However the screening charges are now given by:
\bea
Q_+=\oint e^{i \phi} \\
Q_-=\oint e^{-2i \phi}
\eea
These are clearly not the same as the $\phi$ parts of the $SU(2)_0$ screening charges (\ref{eqn:SU2screen}). Despite this we shall see that the conformal blocks of the two theories are very similar.
\subsection{$c=-2$ structure in $SU(2)_0$}

As was noted in \cite{AlexThesis} the connection with the operators of $c=-2$ seems very close if we arrange the operators of $SU(2)_0$ into the following sets
\bea
\omega&=&\{ j \in Z \}~~~~ h=Z+0 \nonumber \\
\mu&=&\{ j \in \f{4n-1}{2} n \in Z \} ~~~~h=-\f{1}{8}+Z \\
\nu&=&\{ j \in \f{4n+1}{2} n \in Z \} ~~~~h=\f{3}{8}+Z \nonumber
\eea
If we now use the naive tensor product rules resulting from $SU(2)$ i.e. $j \otimes J=|j-J|+ \cdots (|j|+|J|)$ on the above sets we reproduce precisely the known fusion rules for $c=-2$ \cite{Gaberdiel:1998ps} namely:
\bea \label{eqn:fusion}
\nu \otimes \nu &=& \omega \nonumber \\
\nu \otimes \omega &=& \mu \oplus \nu \nonumber \\
\mu \otimes \mu &=& \omega \\
\mu \otimes \nu &=& \omega \nonumber \\
\mu \otimes \omega &=& \mu \oplus \nu \nonumber \\
\omega \otimes \omega &=& 2 \omega \nonumber
\eea
In $c=-2$ these fusion rules were with respect to the classifying $W_3$ algebra whereas the $SU(2)$ ones are with respect to the Kac-Moody algebra. This suggests a deep connection between the $W_3$ algebra of $c=-2$ (actually also an $SU(2)$ triplet) and the Kac-Moody algebra in $SU(2)_0$.

Having seen that the free field representation for the $j=\half$ operator in $SU(2)_0$ is just the dressed $\nu^{\alpha}$ operator in $c=-2$ we can examine the implication of the above fusion rules. From the point of view of $SU(2)_0$ we see that the first rule gives the indecomposable representation found at $h=0,1$ \cite{Caux:1997kq} when two $j=\half$ operators are fused. However the second fusion rule shows that if we fuse this again with another spin $\half$ operator we produce not only spin $\half$ representations but also operators corresponding to the $\mu$ series. In $SU(2)_0$ theory there is an operator with $h=-\f{1}{8}$. However it is the $j=-\half$ operator in the continuous representation of the base $SU(2)$ algebra (i.e $J^a_0$). It seems highly unlikely that the fusion of several discrete representations could yield a continuous one unless the $SU(2)$ structure is lost. The naive tensor product suggests this should be in fact the operator $j=\f{3}{2}$ that has $h=-\f{1}{8} + 2$. It is clear that due to the dressing the identification of operators is not all obvious.\\

In \cite{Kogan:1999hz} fermionic operators were found in the $SU(2)_k$ model at $j=0$. From solving the KZ equations it was found that in the correlation function of four primary operators (all with $j=0, h=0$) it was possible to have logarithmic solutions:
\be
< F_1(z_1) F_2(z_2) F_2(z_3) F_1(z_4) > = \ln \left| \f{z_{12}z_{34}}{z_{13}z_{24}} \right| + 2 \ln  \left| \f{x_{12}x_{34}}{x_{13}x_{24}} \right|
\ee
These fields naturally have a fermionic character as can be seen by permuting the fields.
The $x$ dependence entirely comes from the $SU(2)$ index structure in the theory and so we immediately see that they are \emph{not} in the finite dimensional representations of $SU(2)$.

Ignoring the $x$ dependence it is now clear that these operators are related to the two fermionic vacua $\theta^{\alpha}$ of the $c=-2$ theory. Although these have $h=0$ their OPEs are \cite{Kausch:2000fu}:
\bea
\theta^{\alpha}(z) \theta^{\beta} &\sim& d^{\alpha \beta} (\omega + \ln|z| \Omega) \nonumber \\
\theta^{\alpha}(z) \omega(0) &\sim& -\ln|z| \theta^{\alpha} \\
\omega(z) \omega(0) &\sim& -\ln|z| (2\omega + \ln|z| \Omega) \nonumber
\eea
The $\theta^{\alpha}$ are a fermionic doublet of \emph{primary} operators at $h=0$. The field $\omega$ forms a logarithmic pair with the vacuum $\Omega$:
\bea \label{eqn:omega}
L_0 \omega=\Omega \\
L_0 \Omega=0 \nonumber
\eea

As this same structure was found for general $k$ it suggests that such a structure may be ubiquitous in $SU(2)_k$ for the continuous representations. This is of particular interest as it has been suggested that this $j=-\half$ pre-logarithmic operator \cite{Kogan:1997fd} should play an important role in determining the logarithmic structure of the $\SL$ WZNW model \cite{Nichols:2000mk,Giribet:2001qq}. The $\SL$ model describes string propagation on an $AdS_3$ background which is perhaps the most accessible place to test the AdS/CFT correspondence \cite{Maldacena:1998re}. 

\section{Four point functions in $SU(2)_0$}

As we have see earlier the stress tensor of $OSp(2|2)_{-2}$ is made from commuting $c=2$ and $c=-2$ parts. It was shown in \cite{Ludwig:2000em,Bhaseen:2000mi} that the currents (\ref{eqn:OSpcurrents}) and conformal blocks of $OSp(2|2)_{-2}$ could be expressed in terms of the two sub-theories. All the logarithmic structure originates in the $c=-2$ part of the theory.

As an example we consider the four dimensional representation $[b,q]=[0,\half]$ of $OSp(2|2)_{-2}$ containing the $h=\f{1}{8}$ primary operators (We follow the conventions of \cite{Maassarani:1996jn}). The four operators are:
\be
v_{\pm}=e^{\pm \f{i \phi_1}{\sqrt{2}}} \mu ~~~~ w_{\pm}=e^{\pm \f{i \phi_2}{\sqrt{2}}} \nu^{\pm}
\ee 

We denote by $F_a$ the conformal blocks of the $c=-2$ correlator $\left< \mu \mu\mu\mu \right>$ and by $F_i$ those of the $c=2$ correlator:
\be
\left< e^{+ \f{i \phi_1}{\sqrt{2}}} e^{- \f{i \phi_1}{\sqrt{2}}} e^{+ \f{i \phi_1}{\sqrt{2}}} e^{- \f{i \phi_1}{\sqrt{2}}} \right>
\ee
We then find that the chiral conformal blocks $F_A$ of the four point function in $OSp(2|2)_{-2}$:
\be
\left< v_{+}v_{-}v_{+}v_{-} \right>_{OSp(2|2)_{-2}}
\ee
decompose as a product of the two theories. Namely $F_A=F_a F_i$. We emphasize that this product is at the level of conformal blocks and \emph{not} the full non-chiral correlation functions.

In a similar way one can hope, as was mentioned in \cite{Bhaseen:2000mi}, to express the conformal blocks of $SU(2)_0$ in terms of $c=-2$ and $c=2$ theories. The conformal blocks for the four point function of $j=\half$ and $\nu^{\alpha}$ operators in $SU(2)_0$ and $c=-2$ respectively do appear extremely similar.

In $c=-2$ the four point functions of $\nu^{\alpha}$ operators is:
\bea
\left< \nu^{\alpha,\bar{\alpha}} (z_1,\bar{z}_1) \nu^{\beta \bar{\beta}} (z_2,\bar{z}_2)  \nu^{\gamma,\bar{\gamma}} (z_3,\bar{z}_3)  \nu^{\delta, \bar{\delta}} (z_4,\bar{z}_4) \right> &=& 
|z_{13}z_{24}|^{-\f{3}{2}} |z(1-z)|^{\f{1}{2}} \\
&&\sum_{i,j=1,2}I_i\bar{I}_j (G_a^i \bar{G}_b^j + G_b^i \bar{G}_a^j) \nonumber
\eea
These invariant tensors are $I_1=d^{\alpha \beta}d^{\gamma \delta} ~~~ I_2=d^{\alpha \delta}d^{\beta \gamma}$ and the functions $G_{a,b}^{i,j}$ are now given by:
\bea
G_a^1&=&\f{2E-(1-z)K}{z}~~~~~
G_a^2=\f{-2E+(2-z)K}{(1-z)} \\
G_b^1&=&\f{2 \tilde{E}-(1+z)\tilde{K}}{z} ~~~~~
G_b^2=\f{-2\tilde{E} +z \tilde{K}}{(1-z)} \nonumber
\eea
We can form the expressions for the $SU(2)_0$ blocks $F_{a,b}^{i,j}$ (\ref{eqn:SU2blocks}) from these in the following way:
\bea \label{eqn:su2decomp}
F^1_a&=&\left( z-1 + \f{1}{1-z} \right) G^1_a +(1-z) G^2_a \nonumber \\
F^2_a&=&z G^1_a + \left(-z+\f{1}{z} \right) G^2_a \\
F^1_b&=&-(1-z) G^2_b -\left( z-1 + \f{1}{1-z} \right) G^1_b \nonumber\\
F^2_b&=&- \left( -z + \f{1}{z} \right) G^2_b -z G^1_b \nonumber
\eea
From (\ref{eqn:vertex}) we expect that the $c=2$ free boson part of the operators has conformal dimension $h=0$. We can form several such $h=0$ operators $e^{i \sqrt{2} \alpha (\phi_1 \pm \phi_2)}$ from the compact and non-compact bosons

If we choose $\alpha^2=\f{1}{4}$ then we get precisely the other conformal blocks observed in the correlation functions (\ref{eqn:su2decomp}) for the $c=2$ part namely  $z,\f{1}{z},1-z$ and $\f{1}{1-z}$. It seems that all the logarithmic structure again comes from the $c=-2$ part with slightly different dressing to $OSp(2|2)_{-2}$.

However in $OSp(2|2)_{-2}$ we were able to write some correlators as a direct products rather than as two braided structures. In $SU(2)_0$ this also seems to be the case but the significance of this is unclear as they are not generically valid.

\section{Hamiltonian reduction}
There is another interesting way in which $SU(2)_0$ is related to the $c=-2$ theory. When we do a quantum hamiltonian reduction of $SU(2)_k$ theories normally by imposing the constraint $J^+ \sim 1$ it is well known \cite{Drinfeld:1984qv} that we get to the $c_{k+2,1}$ minimal models. 

The central charge and conformal weights of the reduced theory are given by:
\be
c=c_{k+2,1}=13-6 \left( k+2+\f{1}{k+2} \right) ~~~~~~
h=\f{j(j+1)}{k+2}-j
\ee
Here we follow an elegant realisation of this reduction by Petkova et al. \cite{Furlan:1993mm} that is performed at the level of the correlation functions.

The observation behind the approach is that if we set $x=z$ in the expressions of section (\ref{sec:2and3pt}) we get precisely the correct form for the two and three point functions in the reduced theory with the correct conformal weights.

Surprisingly this simple procedure extends to all the higher order correlators of the theory. Here we shall concentrate on the four point functions of operators in the finite dimensional $SU(2)$ representations. Specifically if one writes the solution to the KZ equation in the form:
\be
\CF(x,z)=\sum_{n=0} (x-z)^n F_n (z)
\ee
then it was shown \cite{Furlan:1993mm} that the lowest component $F_0(z)$, which is the only term surviving in the limit $x \rightarrow z$, obeys the correct differential equation for the primary field $h_{1,s}$ in the $c_{k+2,1}$ model.

Here will shall explicitly calculate some of correlation functions to show how this works in these cases.
For the case of $k=0$ in this way we obtain the $c=-2$ theory (actually it can also be obtained from $k=-\f{3}{2}$ - although not from the finite dimensional representations). The conformal weights of the $c=-2$ fields are given by $\Delta=\f{j(j+1)}{2}-j$, where $j$ is the spin of the $SU(2)_0$ operator. The lowest few are:
\bea
\ba{lllllll}
j~~~~ 0 & \f{1}{2} & 1 & \f{3}{2} & 2 &\f{5}{2} & 3 \nonumber \\
\Delta~~~~ 0 &-\f{1}{8} & 0 & \f{3}{8} & 1 &\f{15}{8} & 3 \nonumber
\ea
\eea
Note in particular that the vacuum logarithmic pair at $h=0$ in the $c=-2$ theory is a direct reduction of the indecomposable representation with $j=0,1$ in the $SU(2)_0$ theory. We also see that the fields $j=2$ and $j=3$ reduce to the fermion fields $\xi^{\pm}$ and W-algebra $W^{a}$ in the triplet model. The extra indicial structure of the $c=-2$ fields \emph{cannot} be explained by this reduction as it is not a part of the normal minimal model. It is however suggestive that to produce the triplet model by reduction an identical index structure should be included in a rational version of the $SU(2)_0$ model.

We have explicitly verified that this simple reduction exactly reproduces all the chiral and non-chiral correlators of \cite{Gaberdiel:1998ps}. Here we give a few examples of correlation functions in the two models:

For example if we take the four point correlator of $j=\half$ operators we have the conformal blocks:
\bea
\CF_1(x,z)&=& z^{\f{1}{4}}(1-z)^{\f{1}{4}} \left\{ \left( -\f{E}{z(1-z)}+\f{K}{z} \right) x + \f{E}{1-z} \right\} \\
\CF_2(x,z)&=& z^{\f{1}{4}}(1-z)^{\f{1}{4}} \left\{ \left( \f{\tilde{E}}{z(1-z)}-\f{\tilde{K}}{1-z} \right) x + \f{\tilde{K}}{1-z} -\f{\tilde{E}}{1-z} \right\} \nonumber
\eea
When we set $x=z$ in these we get the solutions:
\bea
\tilde{\CF}_1(z)&=& z^{\f{1}{4}}(1-z)^{\f{1}{4}} K \\
\tilde{\CF}_2(z)&=& z^{\f{1}{4}}(1-z)^{\f{1}{4}} \tilde{K} \nonumber
\eea
These are precisely the two conformal blocks of the $\left< \mu \mu \mu \mu \right>$ correlator. Moreover in $SU(2)_0$ the non-chiral correlator has the structure:
\be \label{eqn:nonchiralpart}
G(z,\bar{z})=\CF_1(z) \overline{\CF_2}(\bar{z}) + \CF_2(z) \bar{\CF}_1(\bar{z})
\ee
and this reduces to the correct non-chiral correlator in $c=-2$.

The correlation function for the $<\half \half \half \f{3}{2}>$ correlator is:
\bea
\CF_1(x,z)&=& z^{1/4} (1-z)^{1/4} \\
\CF_2(x,z)&=&z^{1/4}(1-z)^{-3/4}\left\{ -1+(1-z) \left( \ln(1-z)-\ln z \right) \right\} +x z^{-3/4} (1-z)^{-3/4} \nonumber
\eea
Setting $x=z$ we get:
\bea
\tilde{\CF}_1(x,z)&=& z^{1/4} (1-z)^{1/4} \\
\tilde{\CF}_2(x,z)&=&z^{1/4}(1-z)^{1/4}\left( \ln(1-z)-\ln z \right)  \nonumber
\eea
These are precisely the conformal blocks of the $\left< \mu \mu \mu \nu \right>$ correlator. The non-chiral correlator in both theories actually only involves the first term $\CF_1$ and $\tilde{\CF}_1$ in $SU(2)_0$ and $c=-2$ respectively.

The correlation function for the $<\half \half \f{3}{2} \f{3}{2}>$ correlator is:
\bea
\CF_1(x,z)&=& z^{1/4}(1-z)^{3/4} F \left(\half, \f{5}{2},1,z \right)+  \nonumber \\
&& x z^{1/4}(1-z)^{-5/4} \f{1}{24} \left\{-z(1-z) F \left( \f{3}{2},\f{3}{2};3;z \right)+(4 z-12)F \left( \f{1}{2},\f{1}{2};2;z \right) \right\} \nonumber \\
\CF_2(x,z)&=& z^{1/4}(1-z)^{3/4} F \left( \half, \f{5}{2};3;1-z \right) + \\
&& xz^{-3/4}(1-z)^{3/4}\f{1}{18} \left\{ (1-z)F \left( \f{3}{2},\f{3}{2};4;1-z \right) +6  F \left( \f{1}{2},\f{1}{2};3;1-z \right) \right\} \nonumber 
\eea
Setting $x=z$ and using standard relations for the hypergeometric functions we get:
\bea
\CF_1(x,z)&=& z^{1/4}(1-z)^{3/4} F \left(\half, \f{5}{2},1,z \right)+  \nonumber \\
&&  z^{5/4}(1-z)^{-5/4} \f{1}{24} \left\{-z(1-z) F \left( \f{3}{2},\f{3}{2};3;z \right)+(4 z-12)F \left( \f{1}{2},\f{1}{2};2;z \right) \right\} \nonumber \\
&=& -\f{1}{48} z^{1/4}(1-z)^{-1/4} \left\{ (18z-48)  F \left( \f{1}{2},\f{1}{2};3;z \right) +(3z^2-2z) F \left( \f{3}{2},\f{3}{2};4;z \right)  \right\} \nonumber \\
&=&  z^{1/4}(1-z)^{-1/4} E\nonumber \\
\CF_2(x,z)&=& z^{1/4}(1-z)^{3/4} F \left( \half, \f{5}{2};3;1-z \right) + \\
&& z^{1/4}(1-z)^{3/4}\f{1}{18} \left\{ (1-z)F \left( \f{3}{2},\f{3}{2};4;1-z \right) +6  F \left( \f{1}{2},\f{1}{2};3;1-z \right) \right\} \nonumber \\
&=& -\f{8}{3} z^{1/4}(1-z)^{-1/4} \left( \tilde{E}-\tilde{K} \right)\nonumber
\eea
These are precisely the conformal blocks of the $\left< \mu \mu \nu \nu \right>$ correlator. In addition both non-chiral correlators have the form of (\ref{eqn:nonchiralpart}).

The correlation function for the $< \f{3}{2} \half \f{3}{2} \f{3}{2}>$ correlator is:
\bea
\CF_1(x,z)&=&  z^{-1/4}(1-z)^{3/4}  +x z^{-5/4}(1-z)^{3/4} \\
\CF_2(x,z)&=& (1-z)^{-5/4} z^{-1/4} - x z^{-5/4}(1-z)^{-5/4}(2z-1) \nonumber
\eea
Setting $x=z$ we get:
\bea
\tilde{\CF}_1(x,z)&=& 2 z^{-1/4}(1-z)^{3/4} \\
\tilde{\CF}_2(x,z)&=& 2 z^{-1/4}(1-z)^{-1/4} \nonumber
\eea
These are precisely the conformal blocks of the $\left< \nu \mu \nu \nu \right>$ correlator.

It was noted in \cite{Petersen:1996xn} that this procedure may become singular in some cases in which case more care would be required.

In the $c=-2$ theory the Kac table is empty and we are forced to extend the representations in order to get a non-trivial theory. In an exactly analogous way we have extended the $SU(2)_0$ beyond the highest weight vector $j=0$ to get a non-trivial theory. Another interesting relation between $c=-2$ and $SU(2)$ (actually ${\mathcal N}=2$ SYM) has been previously noted in the context of Seiberg-Witten theory \cite{Cappelli:1997qf}.

\section{Boundary $SU(2)_0$}

So far we have been discussing correlation functions on the infinite plane where we have no boundaries present. It is well known from the work of Cardy \cite{Cardy:1984bb} how conformal symmetry may also be used to calculate correlation functions in the presence of a boundary. If one chooses conformal boundary conditions then we find that the n point non-chiral correlators in the boundary theory obey the same differential equations as the 2n point chiral bulk ones.

If we consider the theory on the upper half plane, with boundary at  $Im z=0$, then conformal invariance means that the stress tensor obeys $T=\bar{T}$ along the real axis. This allows us to represent $T$ in the lower half plane by $\bar{T}(\bar{z})$. Physically this means that there is no energy flow across the boundary. Here we shall concentrate on the two point correlators of the boundary theory. In the boundary correlator we use the method of images to identify $z_3$ with $\bar{z_2}$ and $z_4$ with $\bar{z_1}$. Then the conformally invariant cross ratio becomes:
\be \label{eqn:boundaryharm}
z=\f{\vert z_1-z_2 \vert^2}{\vert z_1-\bar{z_2} \vert}
\ee
This is clearly always between $0$ and $1$ by simple geometry.

There has recently been discussion of logarithmic CFTs in the presence of a boundary \cite{Moghimi-Araghi:2000cx,Kogan:2000fa,Ishimoto:2001jv,Kawai:2001ur}. In \cite{Kogan:2000fa} it was shown that the logarithmic terms may appear in different limits depending on the boundary conditions present. It was also shown how the boundary states may be defined in principle in LCFTs despite the fact that the Verlinde formula is not valid. In \cite{Kawai:2001ur} the boundary states of the $c=-2$ theory were explicitly constructed.

\subsection{Correlation functions in $SU(2)_0$ in the presence of a boundary}

If we have in addition to the stress tensor $T$, a Kac-Moody current $J^a$, then the boundary conditions on $J^a$ must also be specified. In flat space we have the familiar boundary conditions. We work here in the open string picture:
\be
J^a(w)=\sum_{n \in Z}J^a_n e^{-nw} ~~~~~~ \bar{J}^a(\bar{w})=\sum_{n \in Z}\bar{J}^a_n e^{-n\bar{w}} 
\ee
The boundary is at $w=-\bar{w}$
\bea
\mbox{Neumann} ~~~ (\alpha_n + \tilde{\alpha}_{-n}) | N > =0\\
\mbox{Dirichlet} ~~~ (\alpha_n - \tilde{\alpha}_{-n}) | D > =0
\eea
These are generalised in the WZNW case to:
\bea
\mbox{Neumann} ~~~ (J^a_n + \tilde{J}^a_{-n}) | N > =0\\
\mbox{Dirichlet} ~~~ (J^a_n - \tilde{J}^a_{-n}) | D > =0
\eea
In flat space we may have different conditions in each direction. However in the non-abelian case the boundary condition must be consistent with the $SU(2)$ affine algebra (\ref{eqn:SU2KM}). For instance if we impose Dirichlet conditions in the $+$ and $-$ directions then using:
\be
\left[ (J^+_n - \tilde{J}^+_{-n}),(J^-_n - \tilde{J}^-_{-n}) \right] |B>=2((J^3_n + \tilde{J}^3_{-n})) |B>
\ee
we see that we must have Neumann conditions along the $3$ direction.
In general one looks for boundary conditions of the form:
\be
\left( J^a_n + \tau(\tilde{J}^+_{-n}) \right) |B>=0
\ee
For consistency $\tau$ should be an automorphism of the algebra. We also wish to preserve the Sugawara construction for $T$, in other words $\tau$ should preserve the Killing form, even in the presence of the boundary. We thus consider automorphisms of the form:
\be
\tau(\tilde{J}^+_n)=U^{ab} \tilde{J}^+_n
\ee
with $U^{ab}$ orthogonal. These clearly satisfy the above requirements.

The chiral four point function of spin $\half$ operators in $SU(2)_0$ theory is:
\bea
\left< g_{\epsilon_1} (z_1)  g^{\dagger}_{\epsilon_2} (z_2)  g_{\epsilon_3} (z_3)  g^{\dagger}_{\epsilon_4} (z_4) \right> = 
(z_{13}z_{24})^{-\f{3}{4}} (z(1-z))^{\f{1}{4}} \sum_{i=1,2}I_i F_{a,b}^i
\eea
The invariant tensors are $I_1=\delta_{\epsilon_1,\epsilon_2} \delta_{\epsilon_3,\epsilon_4} ~,~ I_2=\delta_{\epsilon_1,\epsilon_4} \delta_{\epsilon_2,\epsilon_3} $ and the functions $F_{a,b}^{i,j}$ are as given before (\ref{eqn:SU2blocks}). By $SU(2)$ invariance it is clear that non-trivial bulk correlators must have vanishing overall value of $J^3$. In going to the boundary situation there are several choices for how we continue $J^a$ across the boundary:

\begin{itemize}

\item One choice is simply for the currents $J^a$ to behave in a similar way to $T$. With this choice for $Im z <0$ we represent $J^a(z)$ as $\bar{J}^a(\bar{z})$. Then correlators behave as: 
\be
\left< V^+(z_1,\bar{z}_1) \right>_{boundary} = \left< V^+(z_1) V^+(\bar{z}_1) \right>_{bulk} =0 
\ee 
\be
\left< V^+(z_1,\bar{z}_1) V^+(z_2,\bar{z}_2) \right>_{boundary} = \left< V^+(z_1) V^+(z_2) V^+(\bar{z}_2) V^+(\bar{z}_1) \right>_{bulk} =0 
\ee 
This choice clearly preserves the $SU(2)$ invariance of the correlators. Physically these conditions correspond to no current flow through the boundary in other words Neumann boundary conditions.
\item We can also consider twisting the algebra across the boundary by defining for $Im z<0$:
\be
J^+(z)=\bar{J}^-(\bar{z}) ~~~~~~~ J^-(z)=\bar{J}^+(\bar{z})  ~~~~~~~ J^3(z)=-\bar{J}^3(\bar{z})  ~~~~~~~ 
\ee
This preserves both the $SU(2)$ Kac-Moody algebra (\ref{eqn:SU2KM}) and the Sugawara form for $T$ (\ref{eqn:sugawara}). We now get non-vanishing one-point functions:
\be
\left< V^+(z_1,\bar{z}_1) \right>_{boundary} = \left< V^+(z_1) V^-(\bar{z}_1) \right>_{bulk} =|z_1|^{-3/2}
\ee
and the two point function contains logarithmic terms:
\bea
&\left< V^+(z_1,\bar{z}_1) V^+(z_2,\bar{z}_2) \right>_{boundary} = \left< V^+(z_1) V^-(z_2) V^-(\bar{z}_2) V^+(\bar{z}_1) \right>_{bulk} = \nonumber \\
& |z_{13}z_{24}|^{-\f{3}{2}} |z(1-z)|^{\f{1}{2}} \left( \f{K-E}{z} \overline{\f{\tilde{E}}{z}}+ \f{\tilde{E}}{z} \overline{\f{K-E}{z}} \right) 
\eea
where the cross ration $z$ is given by (\ref{eqn:boundaryharm}).
Now at the boundary the current $J^3$ must vanish and so it obeys Dirichlet conditions.

\end{itemize}

\subsection{Logarithmic currents on the boundary with $j=1$ fields}
In \cite{Kogan:2000fa} it was shown that in the presence of appropriate boundary  conditions the boundary operator, which is induced by the field and its image, may or may not be logarithmic. 

For the $j=1$ operators in the $SU(2)_0$ the conformal blocks are \cite{Nichols:2001du}:
\bea \label{spin14pt}
\CF_1(x,z)&=&-\f{1}{2(z-1)}+\f{x}{z}+\f{x^2}{2z(z-1)}  \nonumber  \\
\CF_2(x,z)&=&-\f{1+(z-1)(\ln(1-z)-\ln(z))}{2(z-1)^2}+\f{x(z+(z-1)^2(\ln(1-z)-\ln(z)))}{z(z-1)^2} \nonumber \\ 
    & & + \f{x^2(1-2z+z(z-1)(\ln(1-z)-\ln(z)))}{2z^2(z-1)^2}   \\
\CF_3(x,z)&=&-\f{1-z+\ln(z)}{2(z-1)}+\f{x\ln(z)}{z}+\f{x^2(1-z+z\ln(z))}{2z^2(z-1)}  \nonumber 
\eea
In the non-chiral bulk theory the two chiral theories are combined in the standard way:
\be
G(x,\bar{x},z,\bar{z})=\sum_{i,j=1,2,3} U_{i,j} \CF_i(z) \CF_j(\bar{z})
\ee
In the bulk theory one must impose the constraints of single valuedness and crossing symmetry on this. Firstly imposing single-valuedness everywhere restricts the non-chiral correlator to be:
\bea \label{eq:Gfull}
G(x,\bar{x},z,\bar{z})&=&U_{1,1} \CF_1(x,z) \overline{\CF_1(x,z)} 
+ U_{1,2} \Bigl[ \CF_1(x,z) \overline{\CF_2(x,z)} + \CF_2(x,z) \overline{\CF_1(x,z)} \Bigr] \nonumber \\
& &+ U_{1,3} \Bigl[ \CF_1(x,z) \overline{\CF_3(x,z)} + \CF_3(x,z) \overline{\CF_1(x,z)} \Bigr]
\eea
In order to get a well defined correlator we must also impose the crossing symmetries:
\bea
& &\label{eq:cross1} G(x,\bar{x},z,\bar{z})=G(1-x,1-\bar{x},1-z,1-\bar{z}) \\
& &\label{eq:cross2} G(x,\bar{x},z,\bar{z})=z^{-2h}\bar{z}^{-2h}x^{2j}\bar{x}^{2j} G(\f{1}{x},\f{1}{\bar{x}},\f{1}{z},\f{1}{\bar{z}}) 
\eea

Under $x,z \rightarrow 1-x,1-z$ the solutions transform as:
\be
\CF_1 \rightarrow \CF_1 \quad \CF_2 \rightarrow -\CF_2 \quad \CF_3 \rightarrow \CF_2 + \CF_3
\ee
Under $x,z \rightarrow \f{1}{x},\f{1}{z}$:
\be
\CF_1 \rightarrow \f{z^2}{x^2}\CF_1 \quad \CF_2 \rightarrow \f{z^2}{x^2}(i\pi \CF_1+\CF_2+\CF_3) \quad \CF_3 \rightarrow -\f{z^2}{x^2}\CF_3
\ee
It is easily seen that the only solution obeying both crossing symmetries is $U_{1,2}=U_{1,3}=0$ and so the logarithmic solutions do not contribute to the correlator.  In other words in the bulk theory there are \emph{no} logarithmic terms as $\CF_2$ and $\CF_3$ decouple.

In the presence of a boundary we do not have these conditions and so generically we do expect the logarithmic terms to occur in the full correlator:
\be
A\CF_1+B\CF_2+C\CF_3
\ee
Thus only in the very special case $B=C=0$ would we not observe logarithms in the boundary theory. We thus see that the boundary generically allows us to see logarithmic states that were absent in the bulk theory.

As noted earlier the reduced subspace for the bootstrap may be due to extra singular vectors in the model. If this were true it would mean that these should also decouple in the boundary theory and thus the logarithms could not be observed there either. This is clearly an important point to clarify.

\section{Conclusion}

We have used the Wakimoto free field representation of $\widehat{SU(2)}$ to study some of the structure of the theory at level zero. The stress tensor of $SU(2)_0$ contains within it separate commuting $c=-2$ and $c=2$ sectors. This is precisely the same structure as is known for $OSp(2|2)_{-2}$. This suggests the appearance of a hidden symmetry within both models. The conformal blocks of the two theories seem to be extremely closely related. The $c=-2$ theory is well understood and is known to have logarithmic operators in its spectrum. In both cases all the underlying logarithmic structure comes from this $c=-2$ part. Both these models have $c=0$ and it raises the interesting question as to the universality classes and structures that exist there and how these may be obtained from the underlying current algebras \cite{WorkinProgress}.

Although certain correlators in these models have been studied and they seem suggestive of equivalences they are certainly \emph{not} to be thought of being proven Hilbert space equivalences particularly when the full non-chiral theory is considered. They are certainly more than sheer coincidences although it remains unclear as to how general they really are. To put this on a firmer footing it would be desirable to understand how they arise at the level of the free field representation in the same way as the symplectic fermions in $c=-2$ as this gives the only real hope to understand the full higher point correlators of the theory. Using free fields the situation seems unclear as we often have different screening charges which somehow conspire to yield the same expressions for the correlation function.

When using the auxiliary variable, $x$, to represent to $SU(2)$ group structure we have shown that by the simple procedure of taking $x=z$ one can perform the Hamiltonian reduction of the $SU(2)_0$ theory to $c=-2$ directly at the level of correlation functions. In the examples considered this reduction was valid for the full non-chiral correlators. In the case of the non-compact $SL(2,R)$ WZNW model, which is used to describe string theory on $AdS_3$, the $x$ variable is the coordinate on the two dimensional boundary at spatial infinity. Then we see that the correlation functions on the Hamiltonian reduced theory, in this case Liouville, can be obtained in a simple way from those of $SL(2,R)$. It would be interesting to understand this further.

We have also discussed particular aspects of the $SU(2)_0$ theory in the presence of a boundary. It was found that we will have logarithmic operators with many choices of the boundary conditions. This has potentially interesting consequences for the D-branes in such a model. We examined a case in which although the non-chiral bulk correlator was non-logarithmic (due to the crossing symmetry) the boundary theory can potentially reveal the underlying logarithmic structure.

Clearly many of these points deserve more careful investigation. Another aspect that we have not touched upon is that of modular invariance. It would be very interesting to compare the characters of $SU(2)_0$ representations with those of $c=-2$ and see how they are related \cite{Mukhi:1990bp}. \\

{\bf Acknowledgements}

We would like to thank M. J. Bhaseen and A. Lewis for useful and stimulating discussions. A.N is funded by PPARC studentship number PPA/S/S/1998/02610. I.I.K is partly supported by PPARC rolling grant PPA/G/O/1998/00567 and EC TMR grant HPRN-CT-1999-00161.

\appendix
\section{Operator Product Expansions}
We define the normal ordered products in the standard way by subtracting the singular terms in the OPE and then taking $z \rightarrow w$. In other words:
\be
:A B: (w) \equiv \f{1}{2 \pi i} \oint \f{dz}{z-w} A(z) B(w)
\ee
\subsection{$SU(2)_0$}
\bea
:J^+ J^-: &=&  -2 \beta \xi \eta \gamma -\beta^2 \gamma^2 \nonumber \\
:J^- J^+:&=& -2 \p \xi \eta -2 \xi \p \eta -2 \p \beta \gamma -2 \beta \p \gamma -2 \xi \eta \gamma \beta - \beta^2 \gamma^2 \\
:J^3 J^3:&=& \p \eta \xi + \p \eta \xi + \p \beta \gamma -\p \gamma \beta + 2 \xi \eta \gamma \beta + \gamma^2 \beta^2 \nonumber
\eea
Then:
\be
T=\f{1}{2} \left( \half J^+ J^- + \half J^- J^+ + J^3 J^3 \right) = -\xi \p \eta - \beta \p \gamma
\ee
\subsection{$OSp(2|2)_{-2}$}
\bea
:J J:&=&-2 \p \phi_2 \p \phi_2 \nonumber \\
:H J:&=&:J H:=-2 \p \phi_1 \p \phi_2 \nonumber \\
:H H:&=& -2 \p \phi_1 \p \phi_1 \nonumber \\
:K \hat{K}:&=& 2i \sqrt{2} \p^2 \phi_1 +4 \p \phi_1 \p \phi_1 \nonumber\\
:\hat{K} K:&=& -2i \sqrt{2} \p^2 \phi_1 +4 \p \phi_1 \p \phi_1 \\
: \hat{G}_+ G_-:&=& \f{1}{2} \p \phi_1 \p \phi_1 + \f{1}{2} \p \phi_2 \p \phi_2 + \p \phi_1 \p \phi_2 - \f{i}{\sqrt{2}} \p^2 \phi_1 -\f{i}{\sqrt{2}} \p^2 \phi_2 -2 \xi \p \eta \nonumber \\
:G_- \hat{G}_+:&=& -\f{1}{2} \p \phi_1 \p \phi_1 - \f{1}{2} \p \phi_2 \p \phi_2 - \p \phi_1 \p \phi_2 - \f{i}{\sqrt{2}} \p^2 \phi_1 -\f{i}{\sqrt{2}} \p^2 \phi_2 + 2 \xi \p \eta \nonumber \\
: \hat{G}_- G_+:&=& \f{1}{2} \p \phi_1 \p \phi_1 + \f{1}{2} \p \phi_2 \p \phi_2 - \p \phi_1 \p \phi_2 - \f{i}{\sqrt{2}} \p^2 \phi_1 + \f{i}{\sqrt{2}} \p^2 \phi_2 -2 \xi \p \eta \nonumber \\
:G_+ \hat{G}_-:&=& -\f{1}{2} \p \phi_1 \p \phi_1 - \f{1}{2} \p \phi_2 \p \phi_2 + \p \phi_1 \p \phi_2 - \f{i}{\sqrt{2}} \p^2 \phi_1 + \f{i}{\sqrt{2}} \p^2 \phi_2 + 2 \xi \p \eta \nonumber
\eea
Then:
\bea
T_{OSp(2|2)_{-2}}&=&\f{1}{8} \left( HH-JJ -\half(K \hat{K} + \hat{K} K) + \hat{G}_+ G_- - G_- \hat{G}_+ + \hat{G}_- G_+ - G_+ \hat{G}_- \right) \nonumber \\
&=& - \half \p \phi_1 \p \phi_1 + \half \p \phi_2 \p \phi_2  -\xi \p \eta
\eea
We also have:
\bea
T_{U(1|1)}&=&\f{1}{4} \left( (H+J)(H-J)+ \hat{G}_+ G_- - G_- \hat{G}_+ \right) +\f{1}{8} (J+H)(J+H) \nonumber \\
&=& \f{1}{4} \left( \hat{G}_+ G_- - G_- \hat{G}_+ \right) + \f{3}{8} HH + \f{1}{4} HJ  -\f{1}{8} JJ \nonumber \\
&=& - \half \p \phi_1 \p \phi_1 + \half \p \phi_2 \p \phi_2  -\xi \p \eta
\eea
This explicitly shows the equivalence of these two expressions within the free field representation.
%




\end{document}